\documentstyle[12pt]{article}
\def\bz{{\bar z}}
\def\half{{1\over2}}
\def\del{\partial}

\def\bdel{{\bar\partial}}
\def\bA{{\bar A}}
\def\bX{{\bar X}}
\def\cH{{\cal H}}

\begin{document}
\hsize37truepc\vsize61truepc
\hoffset=-.5truein\voffset=-0.8truein
\setlength{\baselineskip}{17pt plus 1pt minus 1pt}
\setlength{\textheight}{22cm}
%
\begin{titlepage}
\begin{flushright}
TIFR/TH/93-61\\
hep-th/9312190\\
December 1993\\
\end{flushright}
\vspace{2cm}
\begin{center}
{\large\bf The Two-dimensional String as a\\
\vspace{0.3cm}
Topological Field Theory\footnote{Talk given at the Nato Advanced
Research Workshop on {``\it New Developments in String Theory,
Conformal Models and Topological Field Theory''}, Cargese, May 12-21
1993. Based on work done in collaboration with C. Vafa\cite{mukhivafa}
and with D. Ghoshal\cite{dgsm}.}\\}
\vspace{0.8cm}
{\bf Sunil Mukhi}\\
\vspace{0.3cm}
{\it Tata Institute of Fundamental Research}\\
{\it Homi Bhabha Rd, Bombay 400 005, India}
\end{center}
\vspace{2cm}
\begin{center}
ABSTRACT
\end{center}
{\parindent=0pt A certain topological field theory is shown to be
equivalent to the compactified $c=1$ string. This theory is described
in both Kazama-Suzuki coset and Landau-Ginzburg formulations. The
genus-g partition function and genus-0 multi-tachyon correlators of
the $c=1$ string are shown to be calculable in this approach. The KPZ
formulation of non-critical string theory has a natural relation to
this topological model.}
\end{titlepage}
\section{Introduction}

For many years it used to be said that string theory is consistent
only in 26 spacetime dimensions. With the understanding that any CFT
with total central charge $c=26$ is a consistent background for
bosonic string propagation, this statement had to be modified. It
remained true that in some sense the simplest known background for the
bosonic string was the one with 26 flat, noncompact spacetime
dimensions.

The relevance of the concept of ``simplest known background'' stems
from the fact that we do not yet fully understand
background-independent string field theory. In this situation, the
theory in the simplest background may reasonably be expected to
furnish clues about the nature of the full theory. For example,
arbitrary backgrounds in a field theory tend to spontaneously break
some or all of the global symmetries. In such backgrounds we would be
hard pressed to discover the full symmetry structure of the
theory. Thus, we must look for backgrounds which preserve as many
symmetries as possible -- indeed, this is probably the most reasonable
definition of ``simplest background''.

With this in mind, it becomes clear that the noncompact 26-dimensional
background is far from being the simplest one. Indeed, just
compactifying the spacetime on a torus of appropriate shape can
produce\cite{narain} large numbers of Kac-Moody currents, whose
integrals are unbroken symmetry generators. But even better is the
background with just 2 spacetime dimensions, traditionally described
by scalar fields $X(z,\bz)$ and $\phi(z,\bz)$ on the world-sheet. The
unbroken symmetries of this theory correspond to infinitely many
holomorphic currents satisfying the wedge subalgebra of
$W_\infty$\cite{witten,witzwie,klebpoly}. This is the most symmetric
known phase of the bosonic string. More precisely, the maximum number
of unbroken symmetries arise when the field $X(z,\bz)$ is compactified
on a circle of radius ${1\over\sqrt2}$, the so-called self-dual
radius.

The two-dimensional string background (also known as ``$c=1$ string
theory'') has the following energy-momentum tensor for its matter
sector:
\begin{equation}
T(z) = -\half \del X \del X - \half\del\phi\del\phi + \sqrt2
\del^2\phi
\label{emtensor}
\end{equation}
where $X(z,\bz)$ and $\phi(z,\bz)$ are conventionally known as the
``matter'' and ``Liouville'' fields respectively. The presence of a
total derivative term in $T(z)$ corresponds to an extra term in the
worldsheet action proportional to $\int \sqrt{g}R^{(2)}\phi$. This
term renders the functional integral ill-defined, and to stabilise the
theory we must add a cosmological term $\mu
\int\sqrt{g}~{\rm exp}(\alpha\phi)$. This means that Liouville momentum
is not conserved. Every correlation function carries a power of $\mu$
equal to minus the amount by which it violates Liouville momentum
conservation, in multiples of $\sqrt2$.

In particular, it is easy to see that the partition function of the
$c=1$ string in genus $g$ has the behaviour
\begin{equation}
Z_g(\mu) \sim \mu^{2-2g}
\label{zbehaviour}
\end{equation}
Indeed, for the two-dimensional string compactified on the self-dual
radius, the partition function is known explicitly from matrix
models\cite{grosskleb1} and is given by
\begin{equation}
Z_g^{\rm self-dual}(\mu) = \mu^{2-2g}\chi_g,\qquad\quad
\chi_g = {B_{2g}\over 2g(2g-2)}
\label{partfn}
\end{equation}
where $\chi_g$ is the virtual Euler characteristic of the moduli space
of genus-$g$ Riemann surfaces, which is proportional to the Bernoulli
number $B_{2g}$.

It has long been a challenge to reproduce this result from a continuum
formulation of two-dimensional string theory. Other matrix-model
results for this theory, particularly the correlation functions of
``tachyons'' in genus 0, have been re-derived in the continuum only
after considerable difficulty and after resorting to somewhat
arbitrary continuations of the parameters of the
theory\cite{difrk,dotsenko,kitazawa,debswap,jayaram}. One of the
crucial difficulties in continuum calculations is that the theory is
not perturbative in the cosmological constant $\mu$ (as is evident
from Eq.\ref{zbehaviour}).  Another problem is that standard CFT
techniques produce the amplitudes for local vertex operators inserted
at fixed points on the worldsheet, which must then be integrated over
the worldsheet to give the physical amplitudes. In a theory coupled to
gravity, physical answers for correlations of integrated operators
(which are typically quite simple) should not have to depend on first
obtaining the local answer and explicitly performing a complicated
integral.

In what follows I will show that there is a topological field theory
which is equivalent to $c=1$ string theory at the self-dual radius,
and that the manifestly topological formulation permits the
computation of the partition function and amplitudes in any genus, and
at nonzero cosmological constant, without resorting to analytic
continuations. This means that we have a continuum formulation of
this string theory which is apparently as powerful as the very
successful matrix-model description. This discovery may lend itself to
generalization and teach us something about string theory itself
rather than about some specific background, unlike the matrix models which
have so far resisted attempts to go beyond the specific backgrounds
which they represent.

\section{Topological Symmetry of String Backgrounds}

Topological symmetry was thought to be a special property of some
string backgrounds. More recently, it has become clear that it is
present in {\it any} string background in which there is at least one
(possibly anomalous) $U(1)$ current\cite{gato}. The basic idea is that
in conformal gauge, even the bosonic string possesses four chiral
fields of the right spins and statistics to form a twisted (hence
topological) $N=2$ superconformal algebra. In general, this algebra is
generated by bosonic fields $T(z)$ and $J(z)$ of spins 2 and 1
respectively, and fermionic fields $G^+(z)$ and $G^-(z)$ which also
have spins 2 and 1 respectively. They obey the following singular OPE
relations:
\begin{eqnarray}
T(z)T(w) & \sim & {2 T(w)\over (z-w)^2} + {\del T(w)\over (z-w)}
\nonumber \\
T(z) G^\pm(w) & \sim & \half {(3\mp 1)G^\pm(z)\over (z-w)^2} +
{\del G^\pm(w)\over (z-w)} \nonumber \\
T(z)J(w) & \sim & {c^T/3\over (z-w)^3} + {J(w)\over(z-w)^2} +
{\del J(w)\over (z-w)} \nonumber \\
J(z)G^\pm(w) & \sim & \pm {G^\pm(w)\over(z-w)} \nonumber \\
J(z) J(w) & \sim & {c^T/3\over(z-w)^2} \nonumber \\
G^\pm(z) G^\pm(w) & \sim & 0 \nonumber \\
G^+(z) G^-(w) & \sim & {c/3\over (z-w)^3} + {J(w)\over(z-w)^2} +
{T(w)\over (z-w)}
\label{topalg}
\end{eqnarray}

For the $c=1$ string, the bosonic fields $T(z),J(z)$ in the above
algebra arise from the full stress-energy tensor (including ghost
contribution) and the ghost-number current, while the fermionic fields
$G^\pm(z)$ are the BRS current and the antighost. It turns out that
the OPEs among these fields are not actually closed, but on modifying
some of the fields by total derivative terms, one finds a closed
algebra which is precisely the twisted $N=2$ superconformal algebra
above, with some fixed value of the topological central charge $c^T$.

Explicitly, the generators of the twisted $N=2$ algebra are as
follows:
\begin{eqnarray}
T(z) & = & T(z)^{\rm (matter)} + T(z)^{\rm (ghost)} \nonumber \\
G^+(z) & = & c(z) T(z)^{\rm (matter)} + \half :c(z)T(z)^{\rm (ghost)}:
+~ x \del^2 c(z) + y \del(c(z)\del\eta(z)) \nonumber \\
G^-(z) & = & b(z) \nonumber \\
J(z) & = & :c(z)b(z): -~ y \del\eta(z)
\label{topgen}
\end{eqnarray}
where $\del\eta(z)$ is a $U(1)$ current formed out of the fields in the
matter sector of the background. (Here we will work with the case where
the current arises as the derivative of a free scalar field, although
the more general case is obvious.)
If the $\eta$ field has a background
charge $Q_\eta$, then it is easy to check that the constants $x$ and
$y$ must take the following values in order to get a closed algebra:
\begin{eqnarray}
x & = & \half(3 + Q_\eta y) \nonumber \\
y & = & \half(-Q_\eta + \sqrt{Q_\eta^2 - 8})
\label{xy}
\end{eqnarray}
The topological central charge of this algebra is found to be $c^T = 6x$.

Clearly, the structure described above depends on the choice of both a
background and a particular $U(1)$ current of that background.  For
example, if we choose the $c=1$ string, the current can be associated
to either the Liouville scalar $\phi(z, \bz)$ or the matter scalar
$X(z,\bz)$. Thus, we have:
\begin{equation}
\begin{array}{llllll}
\eta = \phi, \qquad & Q_\phi = 2\sqrt2 \qquad & \rightarrow \qquad
& x = -\half, \qquad & y = -\sqrt2, \qquad & c^T = -3  \\
\eta = X, & Q_X = 0 & \rightarrow
& x = {3\over 2}, & y = i\sqrt2, & c^T = 9
\end{array}
\label{examp}
\end{equation}
Another interesting case is the critical bosonic string, in which
one can choose the $U(1)$ current to come from any of
26 free scalars, all non-anomalous, and one obtains $c^T = 9$ in each
case.

The case of interest here will be the non-critical $c=1$ background.
Thus, in principle both the choices of $U(1)$ current above are
available to us. The first choice, in which it is associated to the
Liouville field, may seem more natural since this field is generally
present in noncritical backgrounds\cite{bersh}. However, that choice
has the following disadvantage. If the action is perturbed by the
cosmological operator, as it must be to produce a well-defined
functional integral, the Liouville field satisfies
\begin{equation}
\bdel\del\phi + \mu e^\phi = 0
\label{liouveq}
\end{equation}
Thus, the current $\del\phi$ is no longer holomorphic, and we conclude
that the $\eta$-current in the $N=2$ algebra cannot be identified with
the Liouville field except at zero cosmological constant. No such
restriction applies if we choose instead the $c=1$ matter field (the
``time'' coordinate of the string). Hence we make that choice, and
conclude that $c=1$ string theory is described by a topological $N=2$
algebra with central charge $c^T=9$.

This observation has one more or less immediate consequence. The
various states in the BRS cohomology of the $c=1$ string must be
classified by an $N=2$ topological algebra. It is
well-known\cite{mms,bouwk,lianzuck,witzwie} that there is a
plethora of physical states
in this theory, including tachyons, discrete states of ghost number 1
(the ``remnants'' of transverse gravitons and other tensor excitations
of the string propagating in two dimensions), and discrete states of
ghost number 0 (the so-called ``ground ring'' elements). It can be
argued that the only states among these which are actually primaries
of the topological algebra are the tachyons. This applies at any
compactification radius for the $X$ coordinate, and both the types of
tachyons that generically exist (the special discrete tachyons as well
as the intermediate ones) are always primary. The discrete ``remnant''
states of ghost number 1 are secondaries of the $N=2$ algebra.  On the
other hand, the discrete ground ring states, of ghost number 0, are
both primary and secondary. These states would normally be termed
null, except that (as is familiar\cite{katomat,mms}) the projection
from the chiral algebra to the Fock representation fails to be an
isomorphism, due to the presence of background charges. Hence null
vectors in the module of the chiral algebra can be non-vanishing
Fock-space states.

In our notation, tachyon operators are represented by
\begin{equation}
T_k^\pm = c~{\rm exp}{1\over\sqrt2}(2\mp k)\phi~{\rm exp}{i\over\sqrt2}kX
\label{tachyon}
\end{equation}
where the label $k$ takes all real values in the noncompact case, and
integer values at the self-dual radius. The total conformal dimension
of these operators is of course 0, which is a necessary condition for
them to be in the BRS cohomology. However, one can also consider more
general operators (``off-shell tachyons'') described by
\begin{equation}
T_{k_1, k_2}^\pm =
c~{\rm exp}{1\over\sqrt2}(2\mp k_1)\phi~{\rm exp}{i\over\sqrt2}k_2 X
\label{offshelltachyon}
\end{equation}
Here, we fix the Liouville momentum $k_1$ to be a positive integer,
while the matter momentum $k_2$ takes values in $k_1, k_1-2, \cdots,
-k_1$. These fields have {\em negative} conformal dimension $k_2^2 -
k_1^2$, hence by well-known theorems they cannot be in the BRS
cohomology. In fact, they are BRS exact. But some secondaries of the
topological algebra above these off-shell tachyons are in the BRS
cohomology, and in fact are just the various types of discrete
states.

Thus we see that the presence of a twisted $N=2$ topological algebra
in the $c=1$ string actually explains the variety of physical states
which were discovered over the last few
years\cite{grokleb,berkleb,polyakov}. The tachyons, whose existence
was expected and is rather well-understood, are really the only basic
states of the theory. The two types of discrete states, both of which
seemed a little exotic at first, are in fact nothing but particular
secondaries with respect to the topological algebra. Some details
concerning this phenomenon may be found in Ref.\cite{mukhivafa},
although the full picture probably contains more that has not yet been
worked out.

\section{A Coset Model of the $c=1$ String}

Besides organising the various physical states of the $c=1$ string
elegantly, the topological symmetry described above does something
more fundamental. As it stands, this symmetry looks accidental, in
that there is nothing known about the basics of string theory which
would predict it. But since it exists, one could imagine looking for
alternative formulations of various string backgrounds in which the
topological symmetry is manifest. These formulations might then be
exactly solvable in the way that topological theories often are. For
the special case of the $c=1$ background (at the self-dual radius) we
will argue that there is a manifestly topological model
which fulfils these requirements: it is equivalent to the $c=1$
string, but is ``more solvable'', in the sense that it permits the
computation of amplitudes which are not calculable (as far as we know)
within the conventional KPZ\cite{KPZ} or DDK\cite{DDK} formulations.

The model for which we are searching cannot be predicted from some
general principles, but we do know from the discussion of the previous
section that it must possess a topological symmetry with central
charge $c^T=9$. We will look for this model among a general class of
$N=2$ supersymmetric CFT's, the Landau-Ginzburg and Kazama-Suzuki
models. First we concentrate on the Kazama-Suzuki (KS) description.

The construction of KS models is based on the following facts. Let us
start with an $N=1$ supersymmetric WZW model, based on a group $G$,
and gauge the adjoint action of a subgroup $H$. Then, if and only if
the coset $G/H$ is a K\"ahler manifold, the model so obtained has
$N=2$ supersymmetry\cite{kazamasuzuki}. Applying this idea to the
case where $G=SL(2,R)$ and $H=U(1)$, we find a series of models with
central charge $c=3k/(k-2)$, where $k$ is the level of the $SL(2,R)$
current algebra. After twisting the model so obtained to make it
topological, one finds as usual that the central charge of the theory
becomes zero, but there is a topological central charge $c^T$ which
has the same value as the central charge before twisting. Accordingly,
if we want $c^T=9$ in this class of models, the unique choice is to
take level $k=3$. Curiously, $SL(2,R)$ at level 3 also occurs in the
KPZ description of $c=1$ string theory, and we will see below that
this is not a coincidence -- in fact, the construction described here
will illuminate a few long-standing mysteries of the KPZ approach.

Let us now look at the KS model at level 3 in some detail. The
stress-energy tensor of the supersymmetric $G$ model is, as usual,
\begin{equation}
T^{(G)}(z) = :J^+(z) J^-(z): +~ (J^3(z))^2 - \half(b(z)\del c(z)
+ c(z)\del b(z))
\label{gtensor}
\end{equation}
where $J^+,J^-,J^3$ are the spin-1 $SL(2,R)$ currents, $b,c$ are
spin-$\half$ fermions which serve to supersymmetrise the theory, and
the coefficient of the first term is unity precisely at $k=3$.

One can write the stress-energy tensor of the coset model as the
difference of the (supersymmetric) $G$ and $H$ stress-energy
tensors. However, for our purposes it is more convenient to use an
alternative formulation in which to the $G$ stress-energy we {\em add}
a gauge contribution and then pass to a BRS cohomology on this larger
space\cite{gawkup}. Thus we have
\begin{equation}
T(z) = T^{(G)}(z) + T^{\rm (gauge)}(z)
\label{teetotal}
\end{equation}
We will write this out explicitly below, but first let us see
what the other generators of the $N=2$ algebra look like. Following
Kazama-Suzuki\cite{kazamasuzuki}, we find (at $k=3$)
\begin{eqnarray}
G^+(z) & = & c(z) J^+(z) \nonumber \\
G^-(z) & = & c(z) J^-(z) \nonumber \\
J_{N=2}(z) & = & 3:c(z)b(z): -~2 J^3(z)
\label{gen}
\end{eqnarray}
Here we have labelled the $U(1)$ current of the $N=2$ algebra
explicitly to distinguish it from other $U(1)$ currents in the
problem.

So far, this is an untwisted $N=2$ superconformal algebra, with
central charge $3k/(k-2)$ as mentioned earlier. Now we render it
topological through the twist
\begin{equation}
T(z)\quad\rightarrow\qquad T(z) + \half\del J_{N=2}(z)
\label{twist}
\end{equation}
As a result, the final stress-energy tensor (to be compared with
Eqs.(\ref{gtensor}),(\ref{teetotal})) is
\begin{equation}
T(z) =~ :J^+(z) J^-(z): +~ (J^3(z))^2  - \del J^3(z) - 2 b(z)\del c(z)
+ c(z)\del b(z) + T^{\rm (gauge)}(z)
\label{teetwisted}
\end{equation}
This twist has a rather miraculous consequence. The spins of all the
fields in $T^{(G)}$ have changed, and we now have an $SL(2,R)$ current
multiplet $(J^+,J^3,J^-)$ of spins (2,1,0) respectively. Moreover, the
free fermions $(b,c)$ have changed their spins from $(\half,\half)$ to
$(2,-1)$. The currents are now reminiscent of those in the KPZ
description of $c=1$ string theory, where they described the
gravitational or Liouville sector, while the fermions have become
identical to the usual ghost system of bosonic string theory! The
central charges are found to be $c=27$ for the twisted currents
and of course $c=-26$ for the fermions.

Now we add in the $U(1)$ gauge sector of the theory by making the
gauge choice
\begin{equation}
A_z(z)= \del X(z), \qquad\quad \bA_\bz(\bz)= -\bdel \bX(\bz)
\label{gauge}
\end{equation}
and defining the free scalar field $X(z,\bz)= X(z) - \bX(\bz)$.
Fixing the gauge in this way requires the introduction of a pair of
fermionic ghosts $(B,C)$ of spins $(1,0)$, hence we get
\begin{equation}
T^{\rm (gauge)}(z) = -\half\del X(z) \del X(z) - B(z)\del C(z)
\label{teegauge}
\end{equation}
along with the $U(1)$ BRS charge
\begin{equation}
Q_{U(1)} = \int c(z)(J^3(z) ~- :c(z)b(z): -~ {i\over\sqrt2}\del
X(z))\quad + {\rm c.c}
\label{uonebrs}
\end{equation}

So far, everything except the current-algebra sector has been reduced
to free fields. We now choose to represent the currents also in terms
of free fields, using the Wakimoto representation:
\begin{eqnarray}
J^+(z) & = & :\beta(z)\gamma(z)^2: -~ \sqrt2 \gamma(z)\del\phi(z)
+ 3 \del\gamma(z) \nonumber \\
J^3(z) & = & :\beta(z)\gamma(z): -~{1\over\sqrt2}\del\phi(z) \nonumber \\
J^-(z) & = & \beta(z)
\label{wakrep}
\end{eqnarray}
where $(\beta,\gamma)$ are commuting ghosts and $\phi$ is a free
scalar field. Since we know the spins of the currents (after twisting
of course), it is easy to deduce that the ghosts have spins $(0,1)$
respectively. This means they contribute a total central charge of
+2.  But the total central charge of the twisted current algebra is
27, so we conclude that the field $\phi$ has a central charge $c_\phi
= 25$, which means it must have a background charge
$Q_\phi=2\sqrt2$. Thus, $\phi$ is clearly identical to the Liouville
field in $c=1$ string theory.

To summarise, the full Hilbert space of our topological theory is
\begin{equation}
\begin{array}{cccccccccc}
\cH : & \cH_\phi & \oplus & \cH_X & \oplus & \cH_{b,c} & \oplus
& \cH_{B,C} & \oplus & \cH_{\beta,\gamma} \\
& & & & & & & & & \\
{\rm spins:} & (0) & & (0) & & (2,-1) & & (1,0) & & (0,1) \\
& & & & & & & & & \\
{\rm central~charge:} & 25 & & 1 & & -26 & & -2 & & 2
\end{array}
\label{fullhilb}
\end{equation}
while the physical Hilbert space is obtained from this by taking the
quotient with the two BRS charges
\begin{eqnarray}
G^+ & = & \int cJ^+ = \int c(\beta\gamma^2 - \sqrt2\gamma\del\phi + 3
\del\gamma) \nonumber \\
Q_{U(1)} & = & \int C(\beta\gamma - cb - \del X^-)
\label{twobrs}
\end{eqnarray}
(We have defined $X^\pm = {1\over\sqrt2}(\phi \mp iX)$.)

Note the remarkable fact that the first three Hilbert spaces above are
isomorphic to the Hilbert space of the conventional $c=1$ string
quantised in the DDK formalism. The Liouville field arises from the
Wakimoto representation for the $SL(2,R)$ current algebra, the $c=1$
matter scalar field comes from the $U(1)$ gauge field, and the ghosts
are just the fermions of the original supersymmetric WZW model, after
twisting. The remaining two Hilbert spaces consist of first-order
pairs of the same spins and opposite statistics, so it is quite
reasonable to expect that they cancel out in some sense. Thus, we have
found a very likely candidate for a manifestly topological description
of $c=1$ string theory.

There are two ways to make this connection more convincing. One is to
directly compute the double cohomology of the BRS charges above on the
full Hilbert space. This has been described in detail in
Ref.\cite{mukhivafa} and will not be repeated here. The result turns
out to be isomorphic to the complete cohomology in the DDK formalism,
described in detail in Ref.\cite{witzwie}. Another way is to explore
the relation of this model to the KPZ formalism of $c=1$ string
theory.

\section{Relation to KPZ}

Let us recall the KPZ\cite{KPZ} formulation of non-critical string
theory. One starts with an $SL(2,R)$ current algebra at level $k$, and
twists the Sugawara stress-energy tensor by
\begin{equation}
T_{SL(2,R}\quad\rightarrow\qquad T_{SL(2,R)} - \del J^3(z)
\label{kpztwist}
\end{equation}
The total central charge of this twisted system is
\begin{equation}
c = {3k\over k-2} + 6k
\label{ceekpz}
\end{equation}
Next, one gauges the parabolic subgroup generated by
$J^-(z)$\cite{aleks}. This
introduces a pair of anticommuting ghosts $(B,C)$ of spins $(1,0)$,
along with a BRS charge
\begin{equation}
Q_{KPZ} = \int B(z) (J^-(z) - 1)
\label{kpzbrs}
\end{equation}
Note that in this procedure, the gauge field disappears completely,
unlike in the case where we gauge the subgroup generated by $J^3$.
This is due to the fact that the constraint is first-class\cite{dvvbh}
in the present case.

Now, to the above system (which is collectively supposed to represent
the Liouville, or gravitational, degree of freedom on the worldsheet)
we couple by hand a $c=1$ matter field $X(z,\bz)$ and a pair of
anticommuting ghosts $(b,c)$ of spins $(2,-1)$, and impose the usual
string BRS cohomology via
\begin{equation}
Q_{BRS} = \int c(T_{\rm SL(2,R)} + T_X + \half T_{\rm ghost})
\label{qbrs}
\end{equation}
Summarising, the total Hilbert space is {\em precisely} the same as in
Eq.(\ref{fullhilb}) above, but instead of the two BRS charges in
Eq.(\ref{twobrs}), we have to impose the ones in Eqs.(\ref{kpzbrs}) and
(\ref{qbrs}).

Now to argue the equivalence of our coset model to the KPZ theory, it
only remains to prove that the two double cohomologies on the same
Hilbert space are isomorphic. The proof follows most simply from the
following result\cite{sadov}
\begin{equation}
\beta^{-1} Q_{BRS} = G^+ + \{Q_{U(1)}, * \}
\label{kskpz}
\end{equation}
where ``$*$'' represents some combination of the fields in the theory,
whose detailed form is unimportant. Now on the left hand side, if we
pass to the cohomology of $Q_{KPZ}$ then we can set $\beta=1$
(recalling that in the Wakimoto representation, $J^- = \beta$) and we
are left with $Q_{BRS}$. On the right hand side, if we pass to the
cohomology of $Q_{U(1)}$ then the second term vanishes. Now the
equation just tells us that the operators with respect to which we
want to take the next cohomology are equal, hence the double
cohomologies are equivalent.

This completes the proof that the topological coset model is
equivalent to $c=1$ string theory. In the next sections we will see why
it is a ``more solvable'' formulation of the theory.

\section{Correlation Functions}

In order to show that the $k=3$ coset model is exactly solvable, we
need to investigate the available techniques to solve models of this
kind. Let us first consider a different class of models: the $N=2$
supersymmetric minimal models labelled by a positive integer
$k$. After twisting and coupling to gravity, these models are believed
to represent special points in the moduli space of the $k+1$ matrix
models\cite{kekeli}. In continuum language, these represent the
$(k,1)$ minimal models of central charge $1-6(k-1)^2/k$ coupled to
ordinary gravity, from which the $(k,k')$ models can be obtained by
adding marginal perturbations.

For this class of models, the first powerful technique to be
discovered was the Landau-Ginzburg description\cite{lvw}. In this
approach, one identifies the above theories with the infrared fixed
points of the $N=2$ Landau-Ginzburg (LG) theory with a single
superfield, and superpotential $X^{k+2}$. In this description, it is
known how to explicitly compute correlation functions (after coupling
to gravity) at least in genus 0\cite{dvv,losev}.

The gravitational primaries in the LG theory coupled to gravity are
described by $1, X, X^2, \ldots, X^k$. Let us denote the primary $X^r$
by $U_r$. Gravitational secondaries are obtained by multiplying these
with the usual fields $\sigma_n$ of pure topological gravity. The
selection rules for the correlator $\langle \sigma_{n_1}(U_{r_1})
\ldots\sigma_{n_N}(U_{r_N})\rangle$ are
\begin{equation}
\sum_{i=1}^N ({r_i\over k+2} - 1) + \sum_{i=1}^N n_i =
(2g-2){k+3\over k+2}
\label{oldselrule}
\end{equation}

This picture can be suitably modified for our purposes. For positive
$k$, Landau-Ginzburg theory flows in the infrared to the Kazama-Suzuki
coset $SU(2)_k/U(1)$. Since for many purposes the $SL(2,R)_k/U(1)$
coset is like the coset of $SU(2)$ at level {\em minus} $k$, the model
we have described above should be the infrared fixed point of the LG
theory with $k=-3$, hence with superpotential $X^{-1}$. This requires
some extra work to define carefully, but we will see below that
sphere correlators for $c=1$ string theory can be extracted from this
formalism by making only some very broad assumptions.

To start with, let us restrict to gravitational primaries and set
$k=-3$ in the selection rule above. The resulting equation is
genus-independent, and remarkably simple:
\begin{equation}
\sum_{i=1}^N (r_i + 1) = 0
\label{newselrule}
\end{equation}
This gives us the first clue about the identification of these fields
with the physical fields in the $c=1$ string (at the self-dual
radius). The tachyons in this string theory have discrete momenta,
labelled by integers $k_i$, satisfying momentum conservation in every
genus:
\begin{equation}
\sum_{i=1}^N k_i = 0
\label{tachselrule}
\end{equation}
Thus, we are tempted to make the identification $k_i = r_i + 1$, as a
result of which we would claim that the tachyons in the LG formulation
are gravitational primaries:
\begin{equation}
T_k = X^{k-1}\qquad {\rm (LG)}
\label{tachprim}
\end{equation}

Note that tachyons with all positive and negative integer momenta are
required, so we must allow fields $U_r=X^r$ for all positive and
negative integer values of $r$. It is less clear that {\em all} of
these are gravitational primaries, and for us it will be enough to
treat only $X^r$ for positive $r$ as primaries.

In this framework, it was first noticed by Cecotti and
Vafa\cite{cecvaf} that the correct 4-point function of tachyons in
$c=1$ string theory can be obtained, using techniques that were
developed to deal with polynomial superpotentials. Indeed, it is now
clear that one can obtain the tachyon $N$-point function for all $N$,
using Landau-Ginzburg theory with superpotential $X^{-1}$. This has
been worked out in Ref.\cite{dgsm} using a generalization (to $k=-3$)
of a formalism due to Losev\cite{losev} in which contact terms are
dealt with explicitly, and one finds a recursion formula (on the
sphere) determining correlators of $N$ tachyons in terms of those of
$N-1$ tachyons.

Let us illustrate this first for the 4-point function. Losev's formula
for this is
\begin{eqnarray}
\langle X^{r_1} X^{r_2} X^{r_3} X^{r_4} \rangle_V & = &
{d\over dt_4}\langle X^{r_1} X^{r_2} X^{r_3} \rangle_{V + t_4
X^{r_4}}|_{t_4=0}
\nonumber \\
& & +~ \langle C_V(X^{r_1},X^{r_4}) X^{r_2} X^{r_3} \rangle_V
+ \langle X^{r_1} C_V(X^{r_2},X^{r_4}) X^{r_3} \rangle_V
\nonumber \\
& &
+~ \langle X^{r_1} X^{r_2} C_V(X^{r_3},X^{r_4}) \rangle_V
\label{losevformula}
\end{eqnarray}
where $C_V(X^i, X^j)$ is a contact term between two operators, for
which an expression is known in the case of polynomial LG theory,
where it is described by the so-called Saito pairing. For $k=-3$, it
turns out\cite{dgsm} that the correct contact term is much simpler,
and is given by
\begin{eqnarray}
C_V(X^i, X^j) & = & (i+j)X^{i+j},\qquad (i+j)<0 \nonumber \\
& = & 0,\qquad (i+j)>0
\label{contact}
\end{eqnarray}

The only remaining information required is that the three-point
function is given by\cite{dvv}
\begin{equation}
\langle X^{r_1} X^{r_2} X^{r_3} \rangle_V
= res({X^{r_1} X^{r_2} X^{r_3}\over V'})
\label{threept}
\end{equation}
With all this, the four-point function is easily calculated and one
gets
\begin{equation}
\langle T_{k_1} T_{k_2} T_{k_3} T_{k_4} \rangle_{g=0} =
-\half|k_1 + k_2| -\half|k_1 + k_3| -\half|k_2 + k_3| + 1
\label{tachfourpt}
\end{equation}
which is precisely the tachyon 4-point correlator coming from matrix
models of $c=1$ string theory\cite{matrixpeople} and from
perturbative techniques in the DDK formalism\cite{difrk}.

The same technique can now be applied to the recursive formula for
$N$-point functions, due to Losev and Polyubin in the context of
polynomial LG theories:
\begin{eqnarray}
\langle X^{r_1} X^{r_2} \ldots X^{r_N} \rangle_V & = &
{d\over dt_N}\langle X^{r_1} X^{r_2} \ldots X^{r_{N-1}} \rangle_{V + t_N
X^{r_N}}|_{t_N=0}
\nonumber \\
& & +~ \langle C_V(X^{r_1},X^{r_N}) X^{r_2} \ldots X^{r_{N-1}} \rangle_V
+ \cdots \nonumber \\
& & +~ \langle X^{r_1} X^{r_2} \ldots C_V(X^{r_{N-1}},X^{r_N})
\rangle_V
\label{losevpoly}
\end{eqnarray}

Choosing the kinematic region $k_1 <0$ and $k_2,k_3,\ldots,k_N > 0$,
we find\cite{dgsm}
\begin{equation}
\langle T_{k_1} T_{k_2} \ldots T_{k_N} \rangle_{g=0} =
(k_1 + 1)(k_1+2)\cdots (k_1+N-3)
\label{npoint}
\end{equation}
which is precisely the matrix-model result!

\section{Partition Function}

An alternative method which is in principle even more powerful, is to
identify these theories with the Kazama-Suzuki (KS) coset models based
on $SU(2)_k/U(1)$. It has been shown\cite{wittenks} that
algebraic-geometry techniques can be brought to bear on this problem,
resulting in expressions for arbitrary correlators in arbitrary
genus. These are described as integrals of products of Chern classes
of certain bundles over moduli space.

In the KS formulation, the gravitational primaries
are $1, g_{11}, g_{11}^2, \ldots, g_{11}^k$ (where $g_{ab}$ is
the $SL(2,R)$-valued matrix field of the WZW model). The primary
$g_{11}^r$ is equivalent to the Landau-Ginzburg primary $X^r$ so we
again assign it the label $U_r$. Gravitational secondaries, and the
selection rules for generic correlators, are the same as in the
previous section.

Starting again with genus 0, the 4-point function of gravitational
primaries was explicitly computed for $k>0$ using algebraic-geometry
techniques, in Ref.\cite{wittenks}. This computation is performed by
choosing a section of the relevant bundle and finding its divisor. The
final result is
\begin{eqnarray}
\langle U_{r_1}U_{r_2}U_{r_3}U_{r_4} \rangle_{g=0} & =
& \half( {\rm min}(q_1 + q_2, q_3 + q_4) +
{\rm min}(q_1 + q_3, q_2 + q_4) \nonumber \\
& & + {\rm min}(q_1 + q_4, q_2 + q_3)) - {k+1 \over k+2}
\label{witfourpt}
\end{eqnarray}
where $q_i = r_i/(k+2)$ are the $U(1)$ charges of the fields.

This result can now explicitly be continued to the case of interest to
us.  Inserting $k=-3$ and using the identification in
Eq.(\ref{tachprim}), we easily find the expression in
Eq.(\ref{tachfourpt}). Thus the KS formulation also permits the
calculation of the tachyon four-point function, although in practice
the LG formulation was the simpler one for this case.

Let us finally consider the case of higher genus. Here, two basic
results are known. From Ref.\cite{wittenks}, we know that the
genus-$g$ partition function of the $SU(2)_k/U(1)$ KS model, when
continued to $k=-3$, is precisely the (virtual) Euler characteristic
${\hat\chi}_g$ of the moduli space of genus-$g$ Riemann surfaces with
no punctures.  Now, from the matrix model compactified at the
self-dual radius, we have an explicit result for the genus-$g$
partition function (subject to the assumption that the nonsinglet
sector can be ignored), and the answer is
\begin{equation}
Z_g = {B_{2g}\over 2g(2g-2)}
\label{bernoulli}
\end{equation}
where $B_{2g}$ are the Bernoulli numbers. Remarkably, it has been
shown\cite{harzag,penner,distvaf} that this is just the virtual
Euler characteristic of the moduli space of genus-$g$ Riemann
surfaces. So, our topological formulation of $c=1$ string theory is
powerful enough to reproduce the genus-$g$ partition function,
directly in a continuum approach. It is easy to check that genus-$g$
correlation functions of the cosmological operator ($T_0 = X^{-1} =
g_{11}^{-1}$) are also obtained correctly in our approach.
Correlators of arbitrary (discrete) tachyons in higher genus have yet
to be computed explicitly.

\section{Conclusions}

We have collected enough evidence to show that not only does our
manifestly topological model correctly describe $c=1$ string theory,
but it also allows the explicit computation of correlators which are
either impossible, or very difficult to calculate in any other
continuum formulation (including the conventional conformal-gauge DDK
description). It should be stressed that the correlators described
above are at nonzero cosmological constant, and powers of the
cosmological constant can easily be inserted in the right places in
the above formulae, although we have chosen to omit them for
simplicity of presentation. In contrast, the DDK continuum formulation
requires a perturbative treatment of the cosmological operator, in
addition to certain prescriptions to allow {\em negative} numbers of
insertions of this operator.

The work described here puts $c=1$ string theory on the same footing
as the $c<1$ theory, for which a manifestly topological description
has long been known. However, it leaves some interesting open
questions which need to be addressed in the Lagrangian approaches (LG
or KS). First of all, the role of gravitational descendants in the
topological theory needs to be elucidated. Since such fields are
labelled by two integers, it is tempting to conjecture that they are
related to the discrete states of the $c=1$ string. However, this
identification has so far not been clarified sufficiently. Well-known
properties of $c=1$ string theory like the presence of a ground ring,
and the $W_\infty$ symmetry algebra, appear explicitly in the coset
CFT\cite{mukhivafa}, but are less obvious in the Lagrangian
formulations. Indeed, the important question is not how the KS coset
(described in terms of conformal fields) is related to the DDK $c=1$
string, but rather, how exactly the Lagrangian formulations of this
theory (which are exactly solvable) are related to the CFT
formulation.

\section{Acknowledgements}

I wish to thank the organisers of the Cargese Workshop for their kind
invitation to lecture on this work, and for their generous hospitality
at Cargese. I am grateful to Cumrun Vafa and Debashis Ghoshal for
collaboration on the papers whose results are described here.

\end{document}